# A PROPOSED EXPERIMENT ON THE PROTON DRIVEN PLASMA WAKEFIELD ACCELERATION

G. Xia[1], A. Caldwell[1], K. Lotov[2,3], A. Pukhov[4], R. Assmann[5], F. Zimmermann[5]
1. Max Planck Institute for Physics, 80805, Munich, Germany
2. Budker Institute for Nuclear Physics, 630090, Novosibirsk, Russia
3. Novosibirsk State University, 630090 Novosibirsk, Russia
4. Institute for Theoretische Physics I, Heinrich-Heine-University Duesseldorf, 40225, Germany
5. CERN, Geneva, Switzerland

*Abstract*

Simulations have shown that a high energy, short and intense proton beam can drive a large amplitude plasma wave and accelerate an electron beam to the energy frontier in a single plasma channel. To verify this novel idea, a proof-of-principle demonstration experiment is now being planned. The idea is to use the available high energy proton beams either from the Proton Synchrotron (PS) or the Super Proton Synchrotron (SPS) at CERN, to shoot the beam into a plasma cell and to excite the plasma wakefield. A strong density modulation due to the excited plasma wakefield is produced for a long drive beam and this modulated beam in turn produces a high electric field. The proposed experimental setup is introduced in this paper. The interactions between the plasma and the proton beam are simulated and the results are presented. The compression of an SPS bunch is also discussed.

## INTRODUCTION

Compared to the conventional radio-frequency (RF) metallic cavities used in particle acceleration, using plasma as a medium to transfer energy from the driver beam (laser or electron beam) to the witness beam was shown to achieve very high acceleration gradients [1,2]. In the last decades, more than 3 orders of magnitude higher acceleration gradient than in RF cavities have been demonstrated with plasmas in the laboratory [3,4].

Proton driven plasma wakefield acceleration (PDPWA) has been recently proposed [5]. In this scheme, a high energy (1 TeV), short (bunch length of 100 microns) and intense (bunch intensity of $10^{11}$) proton bunch is shot into a long and uniform plasma cell with plasma density of $6x10^{14}/cm^3$. The space charge of the drive proton beam sets the plasma electrons in motion and excites a large amplitude plasma wakefield. A loaded relativistic electron beam with an energy of 10 GeV as witness beam, surfs the acceleration phase of the wake field and gains very high energy in a single acceleration stage. The particle-in-cell (PIC) simulation results show that 1 TeV proton beam as a driver can bring a bunch of electrons to beyond 600 GeV within 450 m plasma channel. Given the very high beam energy stored at the current proton synchrotrons like Tevatron, SPS and the Large Hadron Collider (LHC) (2 to 3 orders of magnitude higher bunch energy than the SLC beam at SLAC), a new research frontier on the plasma wakefield acceleration is expected to be opened if these huge beam energies could be coupled to the plasma and then to the witness beam. Since PDPWA has this great potential and it could bring the electron beam to the energy frontier (beam energy of Teraelectronvolts, or $10^{12}$ eV) in one stage of acceleration, a proposed experiment to demonstrate the principle of proton driven plasma wakefield acceleration is now being planned at the Max Planck Institute for Physics (MPP) in Munich. The idea is to use the available high energy proton beams either from the PS or the SPS at CERN as drive beams and study the interactions between the proton beams and plasmas. As the injectors for the LHC, the PS and SPS could provide proton beams with maximum momenta of 24 GeV/c and 450 GeV/c respectively. The bunch intensities vary from $10^9$ to $1.6x10^{11}$. If the proton beam is injected into a preformed plasma cell, for example, a uniform Lithium plasma channel, the space charge of the proton beam will pull the plasma electrons onto the axis of beam. Since the mass of the plasma ions is heavier than that of the electrons, the ions are almost immobile. This results in a region near the beam with an excess of plasma electrons. After the beam has passed by, the space charge of excess electrons pushes them back and therefore excites the plasma wakefield. The associated wakefield amplitude can be deduced via the energy variation of the drive beam exiting the plasma cell.

Simulations of interactions between proton beams and plasmas have been done based on the real beam parameters of the PS and SPS [6,7,8]. The results show that the highest wakefields are achieved if the beams are split into micro-bunches by the transverse two-stream instability. The SPS beam can drive a much higher amplitude of the plasma wakefield compared to the field excited by the PS beam. This is largely due to the smaller emittance of the SPS beam. The lower emittance of SPS beam allows the instability to develop before the beam diverges due to the angular spread. In addition, the available tunnel in the current SPS extraction line is around 600 metres long, which far exceeds that of the PS (60 metres maximum). Therefore, it is most likely that the SPS beam will be used in our first demonstration experiment on PDPWA. Next, in this paper, the SPS beam parameters are introduced, followed by the simulation results of interactions between the plasma and the SPS proton beam. The proton bunch length compression is also discussed.

## DRIVE BEAM FROM SPS

The SPS is a part of the injector chain for the LHC. At present, there are two fast extraction lines from the SPS. One is located in the East Area, as shown in Fig.1, which can provide the beam for neutrino physics research at Gran Sasso in Italy and the anti-clockwise beam for LHC through the transfer line TI8. The other fast extraction line and transfer line TI2 from the West Area of SPS can send the clockwise LHC-like beam to the LHC ring. The location of the first demonstration experiment on PDPWA is likely to be in the TT61 tunnel (not shown) in the West Area. The length of tunnel is about 620 meters. The SPS can provide a proton beam with a maximum energy of 450 GeV and an rms bunch length of about 12 cm. The basic beam parameters are listed in Table 1.

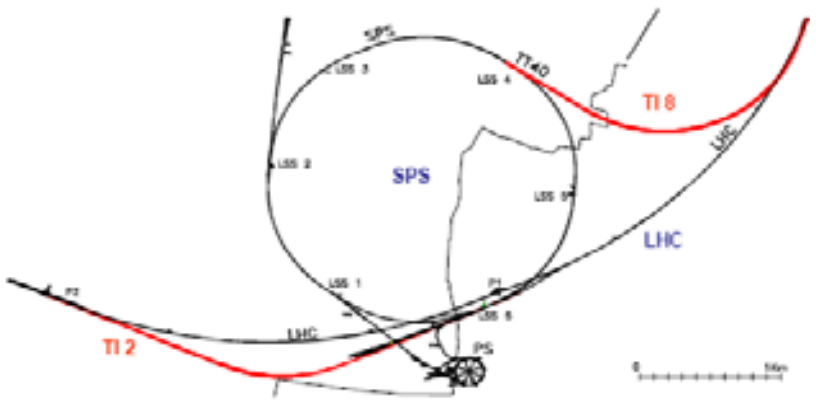

Figure 1: The beam extraction lines from SPS.

Table 1: SPS beam parameters

| SPS beam parameters | |
|---|---|
| Momentum [GeV/c] | 450 |
| Protons/bunch [$10^{11}$] | 1.15 |
| rms longitudinal emittance [eVs] | 0.05 |
| rms bunch length [cm] | 12 |
| Relative rms energy spread [$10^{-4}$] | 2.8 |
| rms transverse normalized emittance [μm] | 3.5 |
| Bunch spacing [ns] | 25 |

To excite a large amplitude plasma wakefield, a short driver is in general required (since the electric field scales as $1/\sigma_z^2$) [9]. However, compression of the SPS bunch length via conventional magnetic compressors from initially 12 cm to the scale of hundreds of microns is difficult due to the high energy of the beam. It requires a lot of RF powers to chirp the beam and big magnets to introduce the dispersive path. To keep the cost of a first experiment as modest as possible, no bunch compression will be implemented. Simulation shows that even without bunch compression, the SPS beam can still excite an interesting plasma wakefield. After the beam is self-modulated by the two-stream instability, the wakefield amplitude (on-axis electric field) can reach several hundred MeV/m. For a plasma density of $10^{15}/cm^3$, simulations indicates that the amplitude of the wakefield can reach more than 200 MeV/m [8]. In addition, if the SPS beam profile can be manipulated, for example, so as to get a 'hard-cut' beam with a steep leading edge, an even higher field is expected. More extensive simulations are in progress to optimize the energy gain achieved in the plasma.

For the first experiment, we expect to see the beam self-modulation effect in the plasma. No electron injection is considered at this stage. A possible beam line layout is shown in Fig. 2. After fine tuning the beam properties like the beam size, beam angular spread, bunch intensity etc, a matched beam (that is, matched to the betatron oscillation amplitude in the plasma) will be shot into a preformed plasma. Since it takes some time for the full self-modulation in the plasma to build up, the length of plasma cell should be of order 5 m at least. After the plasma cell, a beam line with an energy spectrometer can be used to analyze the proton beam energy variation in the plasma. Diagnostic equipment will be employed to characterize the beam properties (beam size, current, emittance, energy etc) with and without the plasma present. The beam dump will eventually stop the spent beam.

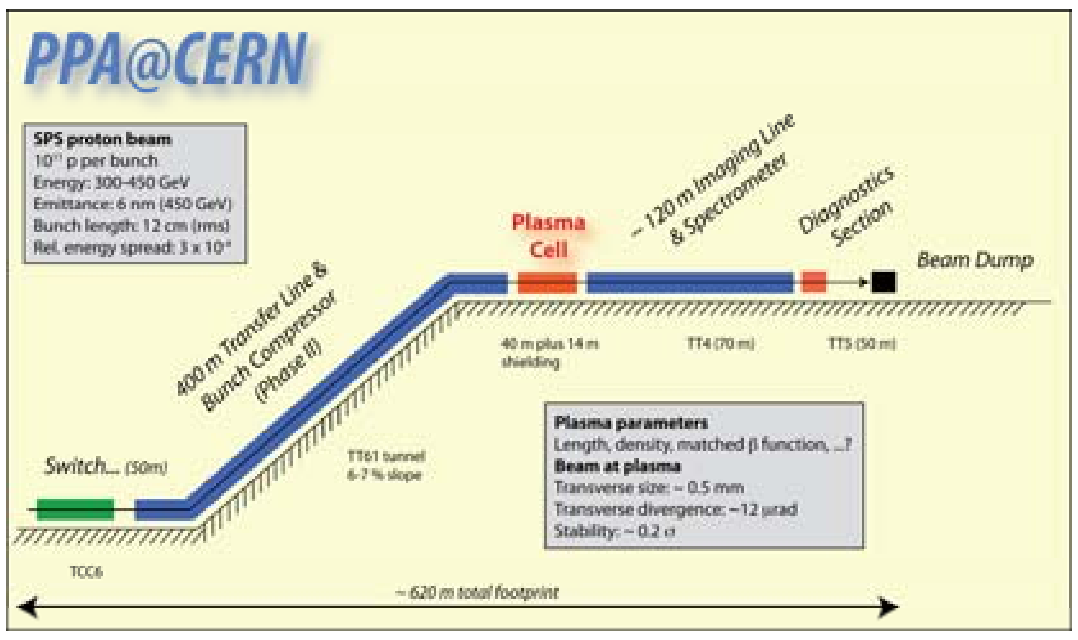

Figure 2: A possible beam line layout for PDPWA experiment.

## SIMULATION RESULTS

Particle-in-cell (PIC) and hybrid codes are usually employed to model the interactions between the plasmas and charged-particle beams. In the PDPWA study, a 2D program LCODE [10,11] and the 3D code VLPL [12] are used to simulate the wakefield excited inside the plasma cell. If the SPS beam propagates directly (without bunch compression) through the plasma, a strong self-modulation occurs, as shown in Fig. 3 (from 3D VLPL code). In this figure (from top to bottom), $E_x$ denotes the longitudinal electric field (beam propagation direction), $E_y$ the radial field, *n* the background plasma density and *n1* the beam density normalized to the background plasma density. It is very much like the self-modulated laser wakefield acceleration (SM-LWFA) concept [13]. The long proton beam with bunch length larger than the plasma wavelength ($\sigma_z >> \lambda_p$) generates a wake within its body, which modulates the bunch itself, leading to an unstable modulation of the whole bunch along the bunch propagation direction. The self-modulation of the long bunch generates a set of ultra-short bunches as beam particles in other regions are pushed to large transverse amplitude. The length of each bunch is around one half of the plasma wavelength $\lambda_p$. These bunches could excite the plasma wake resonantly and the wakefield could be used to accelerate both the protons and also externally injected electrons. Fig. 4 (a) gives the on-axis electric field as a function of travel distance for the SPS beam (without compression) in plasmas of two different densities. It shows that in some conditions, for example, with the smaller beam size and higher plasma density, the wakefield amplitude can reach values beyond 200 MV/m. Fig.4 (b) and (c) show the beam energy spectra for low and high plasma density respectively. It can be seen that within tens of metres in the plasma, some protons in the bunch lose 1 GeV energy and some protons gain 1 GeV energy. More simulation studies are currently ongoing, and optimal parameter regions will be determined.

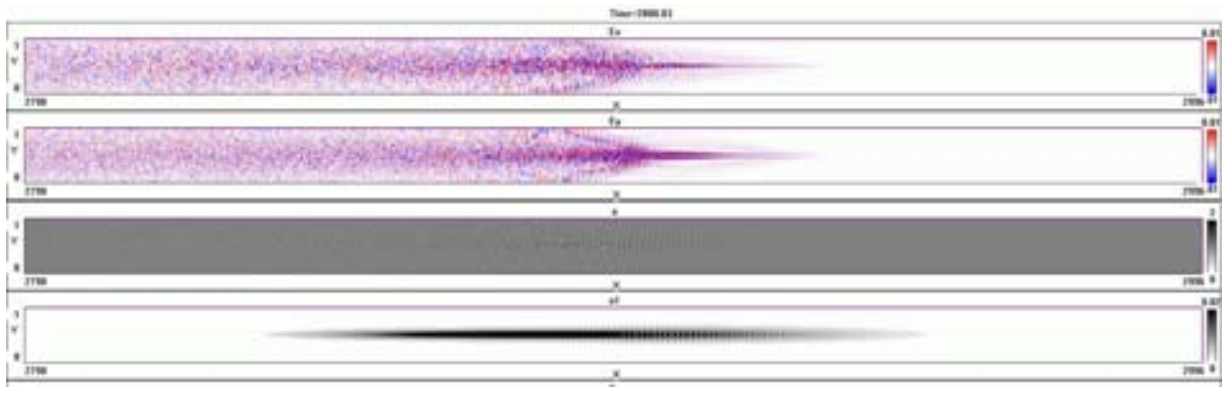

Figure 3: Self-modulation of SPS beam in the plasma.

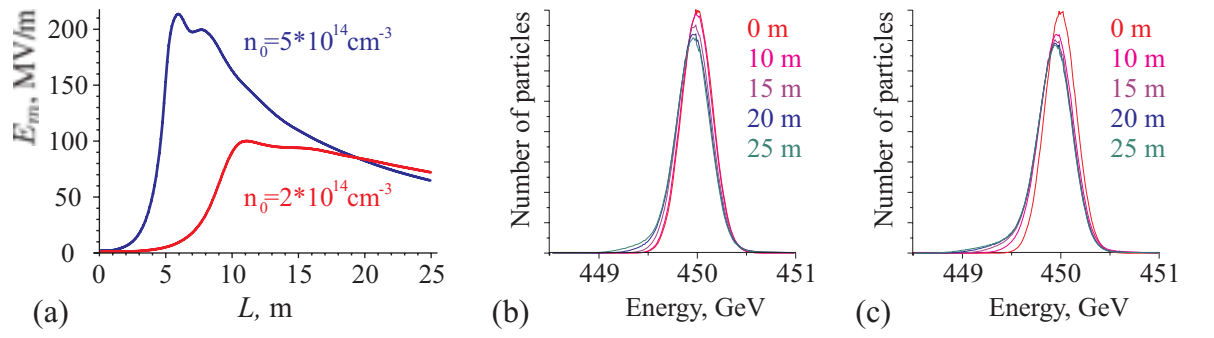

Figure 4: (a) Maximum on-axis electric field as a function of travel distance for the SPS beam in a plasma of two different densities: (b) and (c) beam energy spectra for low and high plasma density cases.

## PROTON BUNCH COMPRESSION

The linear theory of plasma wakefield acceleration indicates that the wakefields' amplitude scales inversely with the square of the bunch length. Therefore, the shorter the driver, the higher the wakefield excited. For the first demonstration experiment, we will not compress the proton bunch from the SPS. However, for future experiments, a bunch compressor which reduces the bunch length to sub-millimetre scale is being considered in order to obtain a more stable and controllable plasma wakefield for electron beam acceleration. However, the current SPS can only provide proton bunches with an rms bunch length of 12 cm or higher. We need to explore how to compress the proton bunch within the available space. A test magnetic bunch compressor has been designed for simulation purposes. It includes RF cavities to provide an energy chirp (position-energy correlation) along the bunch, followed by a dispersive beam line for path modulation (energy-path correlation). The length of the bunch compressor design is around 580 meters. RF sections are assumed to operate at 720 MHz, with a gradient of 25 MV/m. The phase space of the beam before (horizontally flat) and after (sine-like) the bunch compressor is shown in Fig. 5. The simulated bunch length after compression shown in Fig.6 indicates that a 3.6 mm bunch length can be achieved. In this case, the final energy spread is about 2.6x10$^{-2}$. Compared with the initial beam density, the compressed beam density is around 300 times higher than the initial uncompressed beam density. It is expected to excite a large amplitude plasma wakefield if this compressed beam is used as the drive beam.

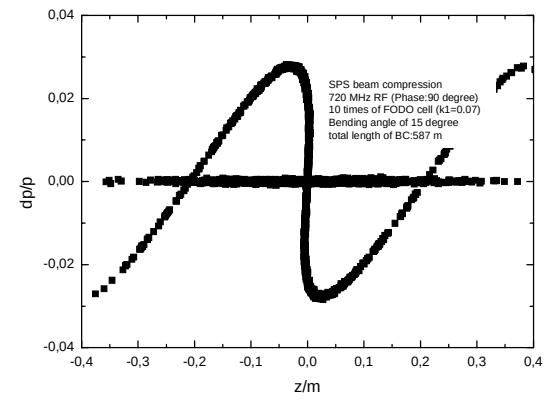

Figure 5: The beam phase space before and after compression.

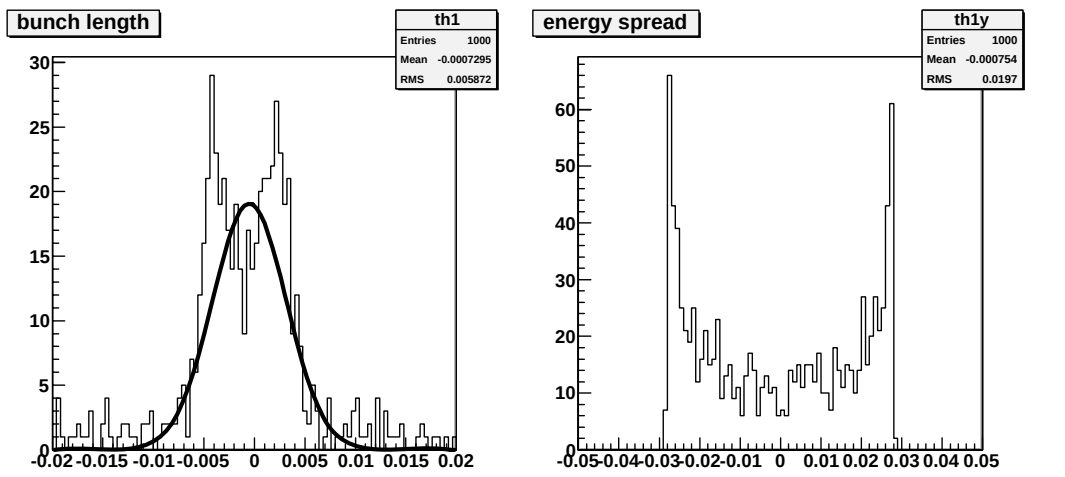

Figure 6: Bunch length and energy spread after bunch compression.

## CONCLUSION

An experimental study of proton driven plasma wakefield acceleration will be proposed at CERN. By using the extracted SPS proton bunch as a drive beam, the interactions between plasmas and protons can be investigated extensively. Simulation results show that for a long proton beam a strong self-modulation occurs in the plasma channel. The associated wakefield amplitude is above 100 MeV/m. A shorter proton bunch produced by a magnetic bunch compressor could excite wakefields of higher amplitude.

## ACKNOWLEDGMENT

The authors would like to thank C. Joshi, B. Bingham, P. Muggli, M. Hogan, C. Huang, W. Lu for stimulating discussions.